\documentclass[twocolumn,aps,showpacs]{revtex4}
\usepackage[utf8]{inputenc}
\usepackage{color}
\usepackage{amsmath}
\usepackage{amssymb}
\usepackage{graphicx}

\makeatletter
\@ifundefined{textcolor}{}
{%
 \definecolor{BLACK}{gray}{0}
 \definecolor{WHITE}{gray}{1}
 \definecolor{RED}{rgb}{1,0,0}
 \definecolor{GREEN}{rgb}{0,1,0}
 \definecolor{BLUE}{rgb}{0,0,1}
 \definecolor{CYAN}{cmyk}{1,0,0,0}
 \definecolor{MAGENTA}{cmyk}{0,1,0,0}
 \definecolor{YELLOW}{cmyk}{0,0,1,0}
}

\usepackage{babel}

\usepackage{babel}

\makeatother

\usepackage{babel}
\begin{document}
\title{Decoherence effects on local quantum Fisher information and quantum coherence in a spin-$1/2$
Ising-$XYZ$ chain }
\author{Hector L.  Carrion$^{1}$,  Onofre Rojas$^{2}$, Cleverson Filgueiras$^{2}$,  Moises Rojas$^{2}$}
\affiliation{$^{1}$School of Science and Technology, Federal University of Rio Grande do Norte, 59078-900, Natal-RN, Brazil,}
\affiliation{$^{2}$Department of Physics, Institute of Natural Science, Federal  University of Lavras, 37200-900,
Lavras-MG, Brazil}
\begin{abstract}
This research explores the effects of decoherence on local quantum Fisher information (LQFI) and quantum coherence dynamics in a spin-1/2 Ising-XYZ chain model with independent reservoirs at zero temperature. Contrasting these effects with those in the spin-1/2 Heisenberg XYZ model reveals intricate interactions among quantum coherence, entanglement, and environmental decoherence in spin systems. Analysis of coherence dynamics highlights differences between the original and hybrid models, showcasing increased entanglement due to Ising interactions alongside reduced coherence from environmental redistribution. LQFI proves more resilient than coherence in specific scenarios, emphasizing decoherence's varying impacts on quantum correlations. This research underscores the complexity of quantum coherence dynamics and the crucial role of environmental factors in shaping quantum correlations, providing insights into entanglement and coherence behavior under environmental influences and guiding future studies in quantum information processing and correlation dynamics.
\end{abstract}
\maketitle

\section{Introduction}

The quantum resource theories \cite{st,st-1,yao} play a crucial role
in the quantum information processing, specifically in the domains of, quantum communication and quantum computation \cite{bra,Bene}. 
These theories identify quantum coherence and entanglement as two fundamental resources. Notably, the most
fascinating nonlocal correlation is the quantum entanglement, an intriguing form of correlation, is widely recognized as a vital physical resource for quantum computation and quantum information.  As the natural candidates for the realization of the quantum coherence, spin chains systems can be used to understand the evolution of the
quantum coherence in its different partitions. Recently, there has been extensive research on the Heisenberg model in the field of condensed matter \cite{wang,wer,wwu}. The influence of quantum coherence on Heisenberg spin models
have been considerable investigated \cite{rada,fan,wei}. Moreover, recent studies reveal a close connection between these resources \cite{adesso}.

The Heisenberg spin diamond chain is a kind of important and interesting model, but the rigorous theoretical treatment is very difficult task due to a non-commutability of spin operators. To overcome problem, a novel class of the simplified versions of the so-called Ising-Heisenberg diamond chain was introduced in Ref. \cite{strec}. Lately, the properties of thermal entanglement have been studied in several Ising-Heisenberg on diamond chain  \cite{moi,cheng,rojas-1,rojas-2}.
In addition, Rojas $et\,al$ \cite{moi-1} discussed the entangled state teleportation through a couple of quantum channels composed of $XXZ$ dimers in an Ising-$XXZ$ diamond chain. More recently, the effects of impurity embedded in an Ising-Heisenberg diamond chain on quantum coherence and quantum teleportation have been reported \cite{moi-2,moi-3,moi-4}.

Recently, a novel measure called local quantum Fisher information $(LQFI)$ \cite{sunkim,yuris} has emerged for quantify 
non-classical correlations through quantum Fisher information $(QFI)$. This powerful quantifier, akin to quantum discord, involves minimizing $QFI$ through a  locally informative observable linked to a specific subsystem. Furthermore, $LQFI$ holds significant promise as a tool to elucidate the influence of quantum correlations, extending beyond entanglement, and enhancing the precision and efficiency of quantum estimation protocols.

On the other hand, it is widely recognized that realistic quantum systems inevitably 
interact with their environment, leading to decoherence, which represents a fundamental challenge in 
quantum information processing \cite{petru,eber,mazzo,sol}. Therefore, the investigation of essential aspects regarding the dynamic behavior of quantum coherence under decoherence has drawn significant interest in recent years \cite{maziero,yang,man,dij,peter,landi}.
In recent years,  extensive research has explored the dissipative effects on the 
dynamics of quantum correlations in Heisenberg spin chains,  analyzing scenarios both in the presence and absence of an external magnetic field \cite{cak,mar,le,tao}. Therefore, it is crucial to continue investigating the study of quantum correlations within  spin models subject to decoherence.

In this paper, based on the previously mentioned developments, we conducted a comprehensive investigation 
into the effects of Ising spins on the dynamics of quantum coherence within the Ising-$XYZ$ spin model. This study was conducted under the 
influence of  environmental decoherence at zero temperature and in the presence affect  of an inhomogeneous magnetic field. The dynamics are modeled using the Lindblad master equation \cite{petru}, from which we derived analytical solutions. These solutions enable us to explore  in detail the concurrence and the behavior of the $l_{1}$-norm of coherence.  Additionally, our analysis allows for the investigation of the dynamics of the local quantum Fisher information.

Our paper is structured as follows. In Sec. II, we introduce the hybrid Ising-$XYZ$
model and derive the Markovian master equation describing the dynamics of the system. In Sec. III, we provide an overview on concurrence, quantum coherence and local quantum Fisher information to quantify the degree of quantum correlations presented in a spin-1/2 Ising-$XYZ$ chain. In Sec. IV, we discuss some of the most interesting results of the dynamics of entanglement, quantum coherence, and local quantum Fisher information of the model studied. Finally, our conclusion is given in Sec. V. 

\section{The model and master equation}

\begin{figure}
\includegraphics[scale=0.5]{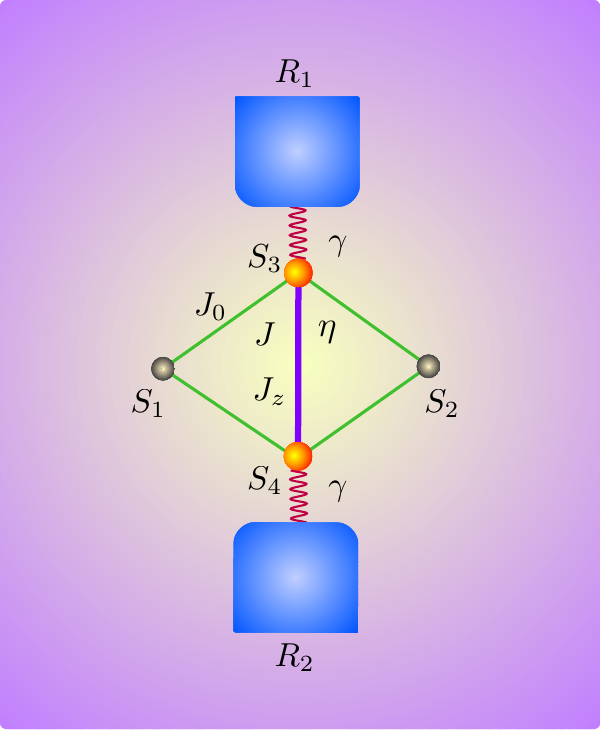} \caption{\label{fig:diamond}(Color online) A schematic representation of structure
of the Ising-$XYZ$ spin chain model with diamond structure connected
with two independent baths $R_{1}$ and $R_{2}$ at zero temperature. The large (orange) circles are the Heisenberg spins, while small (gray) circles are the Ising spins.}
\end{figure}

In this section, we present the Hamiltonian describing the spin-1/2 Ising-$XYZ$
model on a diamond chain subjected to an external inhomogeneous magnetic field. 
The model consists of interstitial Heisenberg spins $(S_{3}^{i},\:S_{4}^{i})$
and Ising spins $(S_{1}^{z},\:S_{2}^{z})$ located in the nodal
site, as illustrated in Fig. \ref{fig:diamond}. The total hamiltonian of
the model may be written as 
\begin{equation}
\begin{array}{cl}
\mathcal{H}= & J\left(1+\eta\right)S_{3}^{x}S_{4}^{x}+J\left(1-\eta\right)S_{3}^{y}S_{4}^{y}+J_{z}S_{3}^{z}S_{4}^{z}\\
 & +J_{0}\left(S_{1}^{z}+S_{2}^{z}\right)\left(S_{3}^{z}+S_{4}^{z}\right)\\
 & +\left(B+b\right)S_{3}^{z}+\left(B-b\right)S_{4}^{z}+\frac{B}{2}\left(S_{1}^{z}+S_{2}^{z}\right).
\end{array}\label{eq:hamil}
\end{equation}
Here, $J$ and $J_{z}$ represent the $XYZ$ interaction
within the Heisenberg dimer, while $\eta$ denotes $XY$-anisotropy parameter.
The nodal-interstitial spins interaction are represented by the Ising-type
exchanges $J_{0}$. Additionally, $B$ denotes the longitudinal uniform magnetic
field in the $z$ direction and $b$ corresponds to the longitudinal nonuniform
magnetic field in the same direction. The matrix form of
the Hamiltonian $\mathcal{H}$ is given by
\[
\mathcal{H}=\left[\begin{array}{cccc}
\mathcal{H}_{1} & \mathbf{O} & \mathbf{O} & \mathbf{O}\\
\mathbf{O} & \mathcal{H}_{0} & \mathcal{\mathbf{O}} & \mathbf{O}\\
\mathbf{\mathbf{O}} & \mathbf{O} & \mathcal{H}_{0} & \mathbf{O}\\
\mathbf{O} & \mathbf{O} & \mathbf{O} & \mathcal{H}_{-1}
\end{array}\right].
\]

Here, we have three distinct blocks: $\mathcal{H}_{1}$ corresponds to the 
$z$-component of spins $\left(S_{1}^{z}=S_{2}^{z}=+1/2\right)$, $\mathcal{H}_{0}$
corresponds to the $z$-component of spins 
$\left(S_{1}^{z}=\pm1/2,\,S_{2}^{z}=\mp1/2\right)$
and $\mathcal{H}_{-1}$ corresponds to the $z$-component of spins
$\left(S_{1}^{z}=S_{2}^{z}=-1/2\right)$.

After diagonalization, the eigenvalues for the $XYZ$ dimer of the Hamiltonian
$\mathcal{H}_{\mu}$ can be obtained as follows 
\begin{align*}
\mathcal{E}_{1,4}= & \frac{B\mu}{2}+\frac{J_{z}}{4}\pm\frac{\Omega}{2},\\
\mathcal{E}_{2,3}= & \frac{B\mu}{2}-\frac{J_{z}}{4}\pm\frac{\omega}{2},
\end{align*}
where $\Omega^{2}=J^{2}\eta^{2}+4\Delta^{2}$, $\omega^{2}=J^{2}+4b^{2}$
, $\Delta=J_{0}\mu+B$ and $\mu=S_{1}^{z}+S_{2}^{z}$. Note that $\mathcal{E}$ depends on $\mu$.

In this paper $\mu$ is the control parameter of the possible values
of the Ising type spins, where $\mu$ assumes the values $\left\{ 0,\pm1\right\} $.
Here $\mu=0$ corresponding the antiferromagnetic case and $\mu=\pm1$
corresponding the ferromagnetic case. We consider the convention $|0\rangle=|\uparrow\rangle$ and $|1\rangle=|\downarrow\rangle$ to indicate the spin-up and spin-down respectively. 

By assuming that the spins $S_{3}$ and $S_{4}$ of the system interacts independently
with the environment, and considering weak system-reservoir
coupling as well as the Born-Markov approximation \cite{petru}, the dynamics
of the dissipative system can be described by master equation, which can be written most generally in the
Lindblad form ($\hbar=1$). Thus, for our specific case, the Lindblad equation takes the following form
\begin{equation}
\frac{d\rho}{dt}=-i\left[\mathcal{H},\rho\right]+\underset{j=3,4}{\sum}\gamma\left[S_{j}^{-}\rho S_{j}^{+}-\frac{1}{2}\left\{ S_{j}^{+}S_{j}^{-},\rho\right\} \right],\label{eq:Li}
\end{equation}
where $\left\{, \right\} $ denotes the anticommutator, and $\gamma$
represents the environmental decoherence rate. Here $\rho$ is the density
operator of the model.
We assume that initially the system's density matrix has the so-called
$X$ form, which is preserved during the evolution according to the
master equation Eq. (\ref{eq:Li}). In the standard basis $\left\{|00\rangle, |01\rangle, |10\rangle, |11\rangle  \right\}$, we have
\[
\rho_{\mu}(t)=\left[\begin{array}{cccc}
\rho_{11}(t) & 0 & 0 & \rho_{14}(t)\\
0 & \rho_{22}(t) & \rho_{23}(t) & 0\\
0 & \rho_{32}(t) & \rho_{33}(t) & 0\\
\rho_{41}(t) & 0 & 0 & \rho_{44}(t)
\end{array}\right],
\]
with $\rho_{41}(t)=\rho_{14}^{\ast}(t)$, $\rho_{32}(t)=\rho_{23}^{\ast}(t)$.

So, the density operator of the model has the following structure
\[
\rho(t)=\left[\begin{array}{cccc}
\rho_{1}(t) & \mathbf{O} & \mathbf{O} & \mathbf{O}\\
\mathbf{O} & \rho_{0}(t) & \mathcal{\mathbf{O}} & \mathbf{O}\\
\mathbf{\mathbf{O}} & \mathbf{O} & \rho_{0}(t) & \mathbf{O}\\
\mathbf{O} & \mathbf{O} & \mathbf{O} & \rho_{-1}(t)
\end{array}\right].
\]

We further suppose that the interacting of the two-qubit Ising-$XYZ$
model with the environment is undergo dephasing
process.

The master equation provided in Eq. (\ref{eq:Li}) is equivalent to
a system of coupled differential equations.  These equations can be solved 
assuming $|\Psi\rangle=\sin\theta|01\rangle+\cos\theta|10\rangle$ as the initial state, resulting in the following expressions
\[
\begin{array}{cl}
\rho_{11}= & \frac{J^{2}\eta^{2}\left[\Omega(1-e^{-2\gamma t})-2\gamma\sin\left(\Omega t\right)e^{-\gamma t}\right]}{4\Omega(\Omega^{2}+\gamma^{2})},\\
\rho_{22}= & \frac{\left[J(2b\sin2\theta-J\cos2\theta)\cos(\omega t)\right]e^{-\gamma t}}{2\omega^{2}}\\
 & -\frac{\left[2\Omega^{2}b^{2}\cos2\theta+\Omega^{2}Jb\sin2\theta-2\Delta^{2}\omega^{2}\right]e^{-\gamma t}}{\Omega^{2}\omega^{2}}\\
 & +\frac{J^{2}\eta^{2}\gamma^{2}\cos\left(\Omega t\right)e^{-\gamma t}}{2\Omega^{2}(\Omega^{2}+\gamma^{2})}+\frac{J^{2}\eta^{2}(1+e^{-2\gamma t})}{4(\Omega^{2}+\gamma^{2})},\\
\rho_{33=} & \frac{\left[J(J\cos2\theta-2b\sin2\theta)\right]\cos(\omega t)e^{-\gamma t}}{2\omega^{2}}\\
 & +\frac{\left[2\Omega^{2}b^{2}\cos2\theta+\Omega^{2}Jb\sin2\theta+2\Delta^{2}\omega^{2}\right]e^{-\gamma t}}{\Omega^{2}\omega^{2}}\\
 & +\frac{J^{2}\eta^{2}\gamma^{2}\cos\left(\Omega t\right)e^{-\gamma t}}{2\Omega^{2}(\Omega^{2}+\gamma^{2})}+\frac{J^{2}\eta^{2}(1+e^{-2\gamma t})}{4(\Omega^{2}+\gamma^{2})},\\
\rho_{44}= & \frac{J^{2}\eta^{2}\gamma\left[\Omega\sin\left(\Omega t\right)-2\gamma\cos\left(\Omega t\right)\right]e^{-\gamma t}}{2\Omega^{2}\left(\Omega^{2}+\gamma^{2}\right)}\\
 & -\frac{4\Delta^{2}e^{-\gamma t}}{\Omega^{2}}+\frac{-J^{2}\eta^{2}e^{-2\gamma t}+\Omega^{2}+12\Delta^{2}+4\gamma^{2}}{4\left(\Omega^{2}+\gamma^{2}\right)},\\
\rho_{23}= & \frac{-\left(J\cos2\theta-2b\sin2\theta\right)\left[i\omega\sin(\omega t)+2b\cos(\omega t)\right]e^{-\gamma t}}{2\omega^{2}}\\
 & +\frac{J\left(J\sin2\theta+2b\cos2\theta\right)e^{-\gamma t}}{2\omega^{2}},\\
\rho_{14}= & \frac{J\eta\gamma\left[\left(i\gamma+2\Delta\right)\Omega\sin\left(\Omega t\right)+\left(i\Omega^{2}-2\gamma\Delta\right)\cos\left(\Omega t\right)\right]e^{-\gamma t}}{2\Omega^{2}\left(\Omega^{2}+\gamma^{2}\right)}\\
 & +\frac{J\eta\left[2\Delta\left(\Omega^{2}+\gamma^{2}\right)e^{-\gamma t}-\left(i\gamma+2\Delta\right)\Omega^{2}\right]}{2\Omega^{2}\left(\Omega^{2}+\gamma^{2}\right)}.
\end{array}
\]

\section{Dynamics of Quantum Correlations}

In this section, we will focus on studying the dynamics of the entanglement, 
quantum coherence and local quantum Fisher information. Specifically,  we aim 
to investigate the entanglement dynamics of anisotropic Heisenberg qubits on 
the Ising-$XYZ$ model. To accomplish this, we will utilize the 
concurrence $\mathcal{C}$ as a measure of entanglement, as defined by \cite{hill, woo}
\begin{equation}
\mathcal{C}(\rho)=\mathrm{max}\{\sqrt{\lambda_{1}}-\sqrt{\lambda_{2}}-\sqrt{\lambda_{3}}-\sqrt{\lambda_{4}},0\}\;,\label{eq:Concurrence}
\end{equation}
where $\lambda_{i}$ $\left(i=1,2,3,4\right)$ are the eigenvalues
in a decreasing order of the operator $R$, which is given by 
\begin{equation}
R=\rho\cdot\left(\sigma^{y}\otimes\sigma^{y}\right)\cdot\rho^{*}\cdot\left(\sigma^{y}\otimes\sigma^{y}\right)\;.\label{eq:R}
\end{equation}
$\rho^{*}$ denotes the complex conjugate of matrix $\rho$. In our
model, the concurrence for a state of the form given by Eq. (\ref{eq:Concurrence}) is
given by
\begin{equation}
\mathcal{C}=2\mathrm{max}\{|\rho_{23}|-\sqrt{\rho_{11}\rho_{44}},|\rho_{14}|-\sqrt{\rho_{22}\rho_{33}},0\}\;.\label{eq:conc-def}
\end{equation}

On the other hand, the quantum coherence is a useful resource for
the quantum information processing task. Here, we will employ the
$l_{1}$-norm\cite{baum} measure, defined as 
\[
C_{l_{1}}(\rho)=\underset{i\neq j}{\sum}|\rho_{ij}|\;.
\]
The corresponding $l_{1}$-norm of the quantum coherence of the impure
dimer described by the reduced density operator, ${\rho}(t)$,
is given by 
\begin{equation}
C_{l_{1}}=2|\rho_{2,3}|+2|\rho_{1,4}|\;.\label{eq:Cl1}
\end{equation}
It is worth noting that quantum coherence is a basis-dependent concept. To analyze $\mathcal{C}_{l_{1}}$, we perform a unitary transformation on the density matrix of the bipartite system $\rho_{AB}$. The transformed density matrix is given by  $\widetilde\rho_{AB}=\widetilde U\rho_{AB}\widetilde U^{\dagger}$, where  $\widetilde U=U\otimes U$. The unitary matrix $U$ is defined as
\begin{equation}\label{u1}
U=\left[\begin{array}{cc}
\cos\phi & -e^{i\varphi}\sin\phi\\
e^{-i\varphi}\sin\phi & \cos{\phi} 
\end{array}\right].
\end{equation}

The quantum Fisher information is essential in quantum estimation theory, serving as the quantum equivalent of classical Fisher information. This quantifier provides us with a profound understanding of how quantum correlations play a crucial role in determining metrological precision. Specifically, when $\rho_{\theta}=e^{-i\theta H} \rho e^{i\theta H}$, where $H$ is a fixed Hermitian operator on the system $\rho$, the quantum Fisher information becomes independent of the estimated parameter $\theta$ \cite{luo,Li}.
\begin{equation}
\mathcal{F}(\rho,H)=\frac{1}{2}\sum_{i\neq j}\frac{(p_{i}-p_{j})^2}{p_{i}+p_{j}}|\left<\psi_{i}|H |\psi_{j}\right>|^{2},
\end{equation}
where $p_{i}(p_{j})$ and $|\psi_{i} \rangle$ ($|\psi_{j} \rangle$) are the eigenvalues and eigenvectors, respectively, of the density matrix $\rho$. Now, assuming that the dynamics is governed by local Hamiltonian $H=H_{A}\otimes\mathbb{I}$, with $H_{A}$ defined as $H_{A}=\boldsymbol{\sigma}\cdot \mathbf{r}$, where $|\mathbf{r}|=1$ and $\boldsymbol{\sigma}$ is the Pauli spin vectors, we can derive the quantification of quantum correlations in terms of local quantum Fisher information ($LQFI$) as follows
\begin{equation}
\mathcal{Q}_{\mathcal{F}}=1-\lambda_{max}^{\mathcal{M}},  
\end{equation}
where $\lambda_{max}^{\mathcal{M}} $ denotes the largest eigenvalue of the symmetric matrix $\mathcal{M}$ with elements
\begin{equation}
\mathcal{M}_{lk}=\sum_{i\neq j}\frac{2p_{i}p_{j}}{p_{i}+p_{j}} \left<\psi_{i}|\sigma_{l}\otimes\mathbb{I} |\psi_{j}\right>\left<\psi_{j}|\sigma_{k}\otimes\mathbb{I}|\psi_{i}\right>. 
\end{equation}

\begin{figure*}
\includegraphics[scale=0.35]{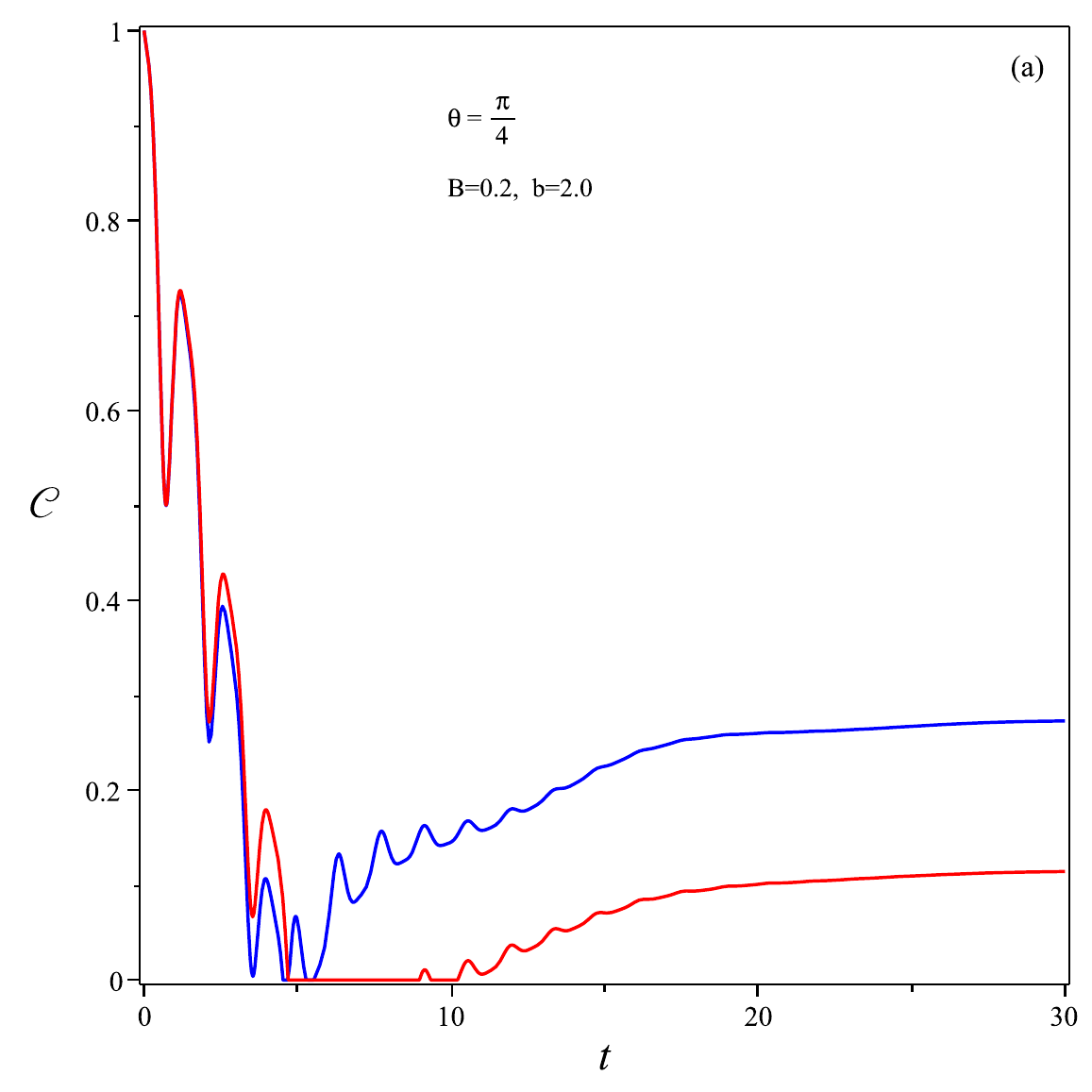}\includegraphics[scale=0.35]{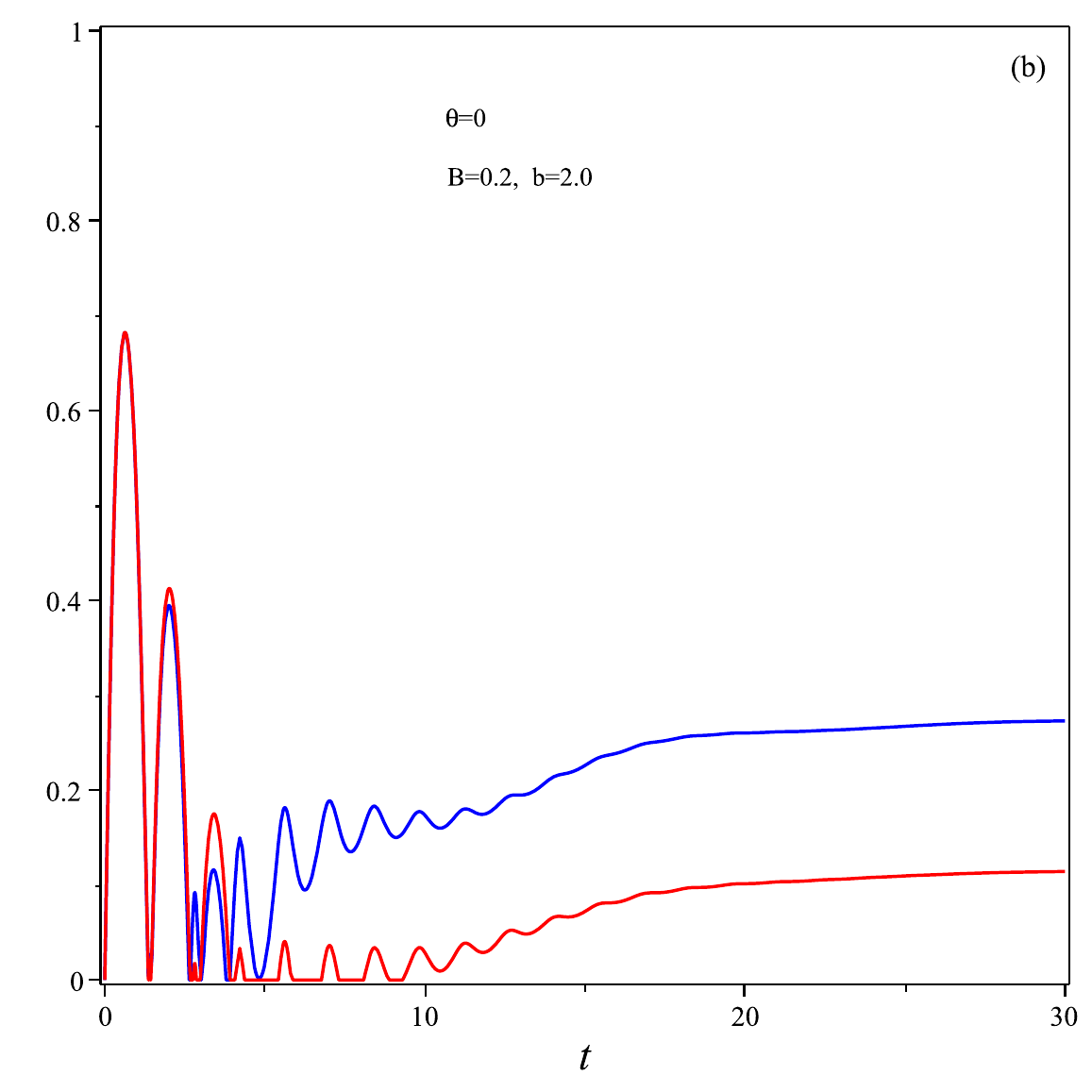}

\caption{\label{fig:C1vst}(Color online) The concurrence $\mathcal{C}$ as
a function of time $t$, with $J=2.0$, $J_{0}=1.0$, $\gamma=0.2$, $\eta=0.2$.
Here, the blue curve corresponds to two-qubit Heisenberg $XYZ$ model
and red curve corresponds to two-qubit Ising-$XYZ$ model. (a) $\theta=\frac{\pi}{4}$.
(b) $\theta=0$.}
\end{figure*}


\section{Results and discussion}

In this section, we outline the main results regarding the dynamical of concurrence, quantum coherence, and local quantum Fisher information for the spin-$1/2$ Ising-$XYZ$ chain.

\subsection{Concurrence}

In the following, we compare the time evolution of entanglement between the original spin-1/2 Heisenberg $XYZ$ model  \cite{tao} and the 
hybrid spin-1/2 Ising-$XYX$ model. In this case, the two-qubits system is initially prepared in the
state $|\Psi\rangle=\sin\theta|01\rangle+\cos\theta|10\rangle$. In Fig. \ref{fig:C1vst}, the 
concurrence $\mathcal{C}$, as a function of the time $t$ is depicted for $J=2.0$, $J_{0}=1.0$, $\gamma=0.2$, $\eta=0.2$, $B=0.2$, $b=2.0$, and for various values of the parameter $\theta$. Here, the red curve represents the concurrence for the Heisenberg model, while the blue curve represents the concurrence for the Ising-$XYZ$ chain. 
Thus, in Fig. \ref{fig:C1vst}(a), we display the concurrence  $\mathcal{C}$ for initial state $|\Psi\rangle=\frac{1}{\sqrt{2}}(|01\rangle+|10\rangle)$, 
which is maximally entangled $\left(\theta=\pi/4\right)$. We can observe that the concurrence for the Ising-$XYZ$ model is slightly more robust than in 
the original model until reaching the first entanglement sudden death (ESD), after which there is a reversal in entanglement efficiency. In this region, entanglement is notably more robust in the original model than in the Ising-$XYZ$ spin chain model.
On the other hand, we are also interested in the behavior of concurrence when the initial state is completely 
non-entanglement. Thus, in Fig.  \ref{fig:C1vst}(b), we depict the concurrence for the initially unentangled state $|\Psi\rangle=|01\rangle$ $\left(\theta=0\right)$.
\begin{figure*}
\includegraphics[scale=0.35]{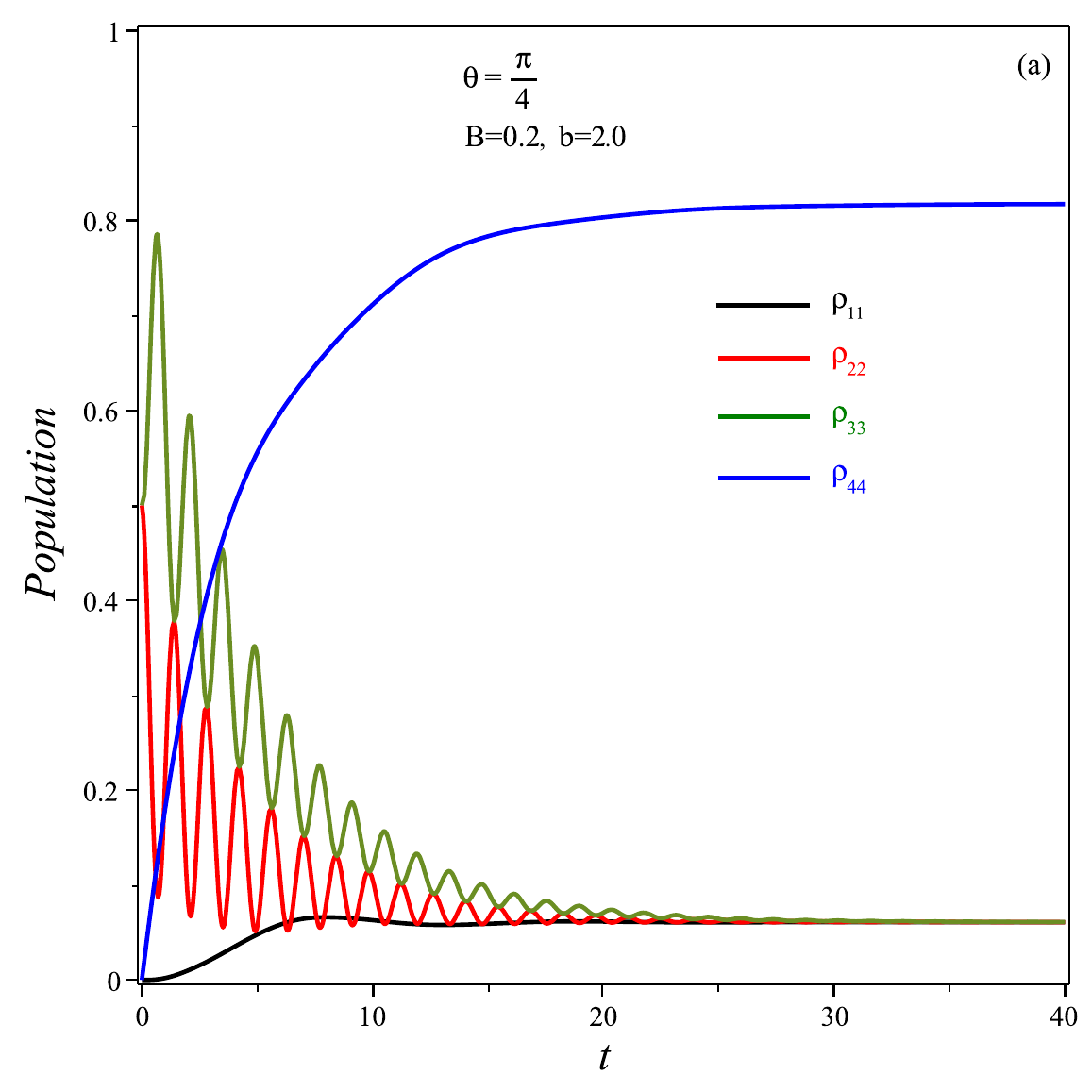}
\includegraphics[scale=0.35]{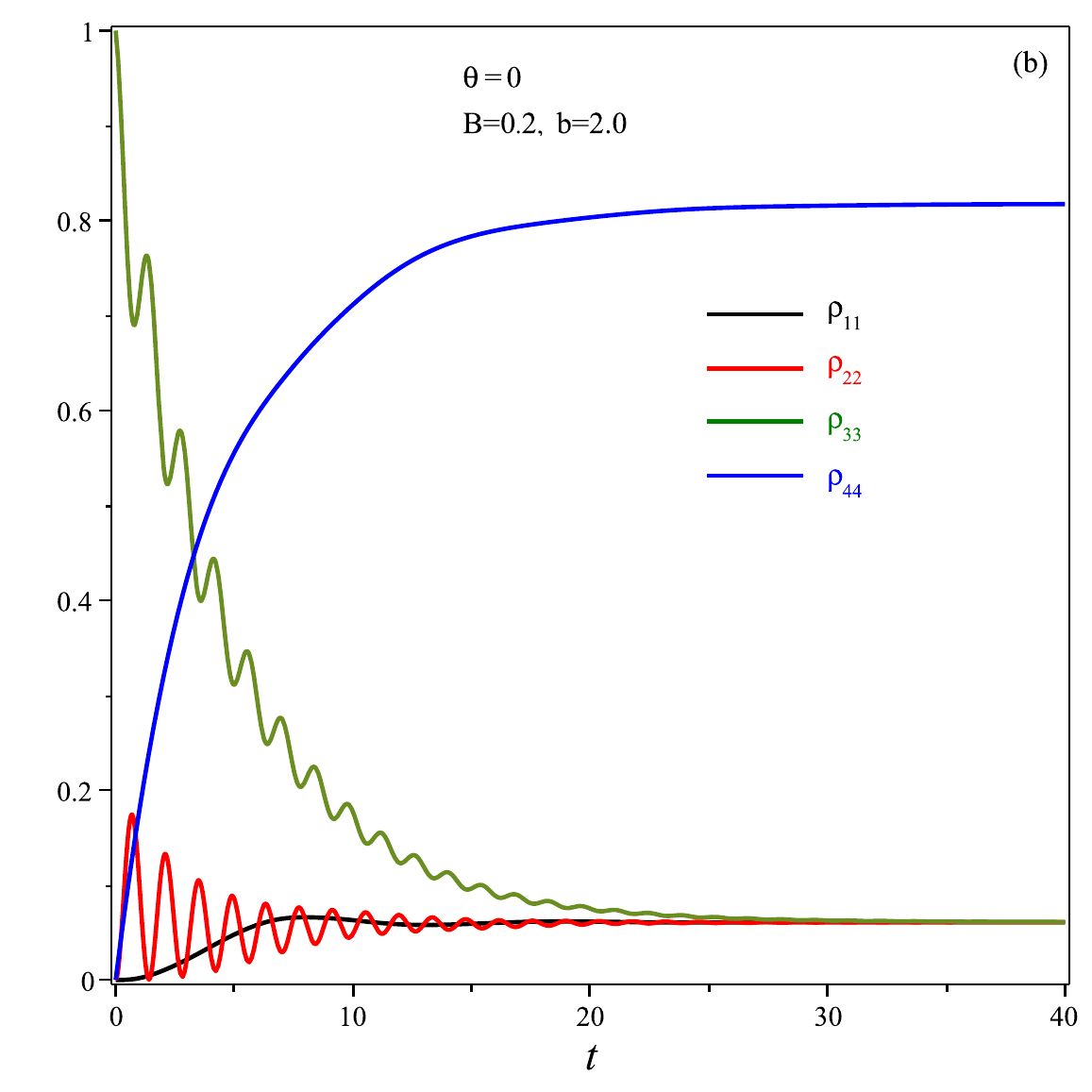}

\caption{\label{fig:popu}(Color online) The population as a
function of time $t$ for Ising-$XYZ$ model. (a) $\theta=\frac{\pi}{4}$. (b)
$\theta=0$. The parameters are chosen as $J=2.0$, $J_{0}=1.0$, $\gamma=0.2$,
$\eta=0.2$.}
\end{figure*}
It is easy to see that we have three well-defined regions of concurrence 
behavior. In the initial sector, one can observe that initially the concurrence is null for both models and for short time intervals, the concurrence of the original model and hybrid models is similar. However, in this region, the hybrid model (red curve) is slightly more robust than the original 
model (blue curve). In the second sector, there is a significant change in the concurrence behavior.  In this region, both models experience entanglement sudden death (ESD) and entanglement sudden birth (ESB) in a short time \cite{yu,ta}. Finally, in the last sector, the entanglement of the
models grows progressively, with the predominance of the Heisenberg model's over  the Ising-$XYZ$ model. Ultimately, the concurrence for each model reaches a steady state after exhibiting some oscillatory behavior.
\begin{figure}
\includegraphics[scale=0.38]{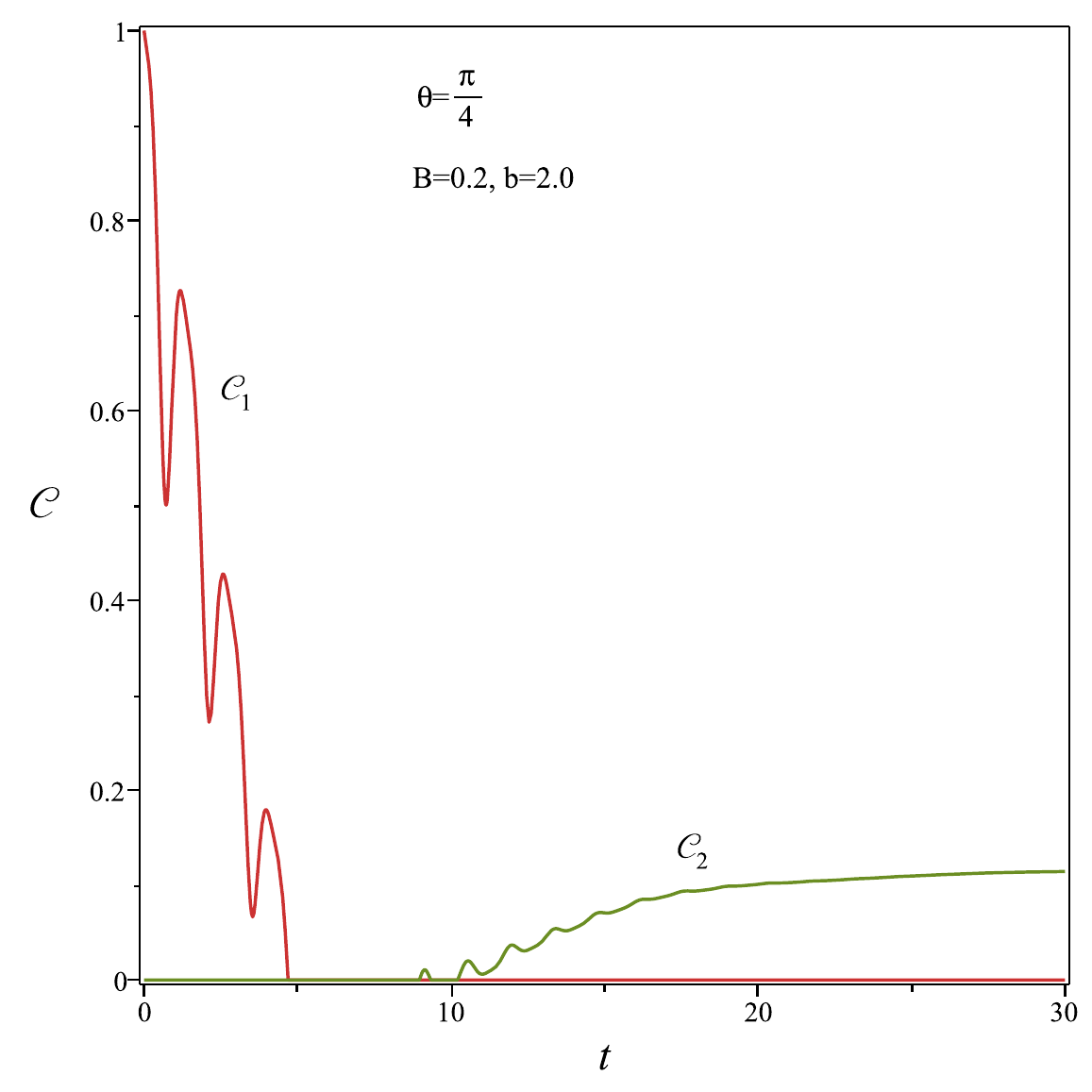}

\caption{\label{fig:concurrence}(Color online) Concurrence $\mathcal{C}_{1}$ and $\mathcal{C}_{2}$ plotted
against time $t$ for the Ising-$XYZ$ chain with $B=0.2$, $b=2.0$.  The other parameters are set to $J=2.0$, $J_{0}=1.0$, $B=0.2$, $\gamma=0.2$,
$\eta=0.2$ and $\theta=\pi/4$.}
\end{figure}

In the following, we study the population dynamic of the hybrid Ising-$XYZ$ diamond chain. Fig. \ref{fig:popu} illustrates 
the dynamics of population $\rho_{ii}$ $(i=1,2,3,4)$ for the initial 
state $|\Psi\rangle=\sin\theta|01\rangle+\cos\theta|10\rangle$, with specific emphasis on states corresponding for $\theta=\pi/4$ and $\theta=0$. In both cases we fixed the parameters $J=2.0$, $J_{0}=1.0$, $\gamma=0.2$, $\eta=0.2$. 
Specifically, Fig. \ref{fig:popu}(a) show the evolution of population dynamics for the entanglement state $(\theta=\frac{\pi}{4})$. On a short-time scale,  
the population $\rho_{22}$ (red curve) and $\rho_{33}$ (green curve) corresponding to Bell state exhibit fast oscillations.  Notably, the 
populations of $\rho_{11}$ (black curve) and $\rho_{44}$ (blue curve), which were initially zero, experience a rapid increase for $\rho_{44}$, while 
the population associated  with $\rho_{11}$ exhibits a 
slower increment. Fig. \ref{fig:popu}(b) corresponds to the case when 
the system is initially prepared in the unentagled  
state $|\Psi\rangle=|10\rangle$ $(\theta=0)$. Similarly to case in Fig. \ref{fig:popu}(a), it is 
observed here that the initially zero populations  of $\rho_{44}$ increase rapidly, while 
the population corresponding to $\rho_{11}$ shows a slower increment. On the other hand, it 
is evident that the system starts in the state $\rho_{33}$. As time progresses, the 
state $\rho_{22}$ and $\rho_{33}$ undergo a series of oscillations, culminating in the 
emergence of quantum entanglement (see Fig. \ref{fig:C1vst}(b)) due to interaction with 
the environment.
\begin{figure}
\includegraphics[scale=0.5]{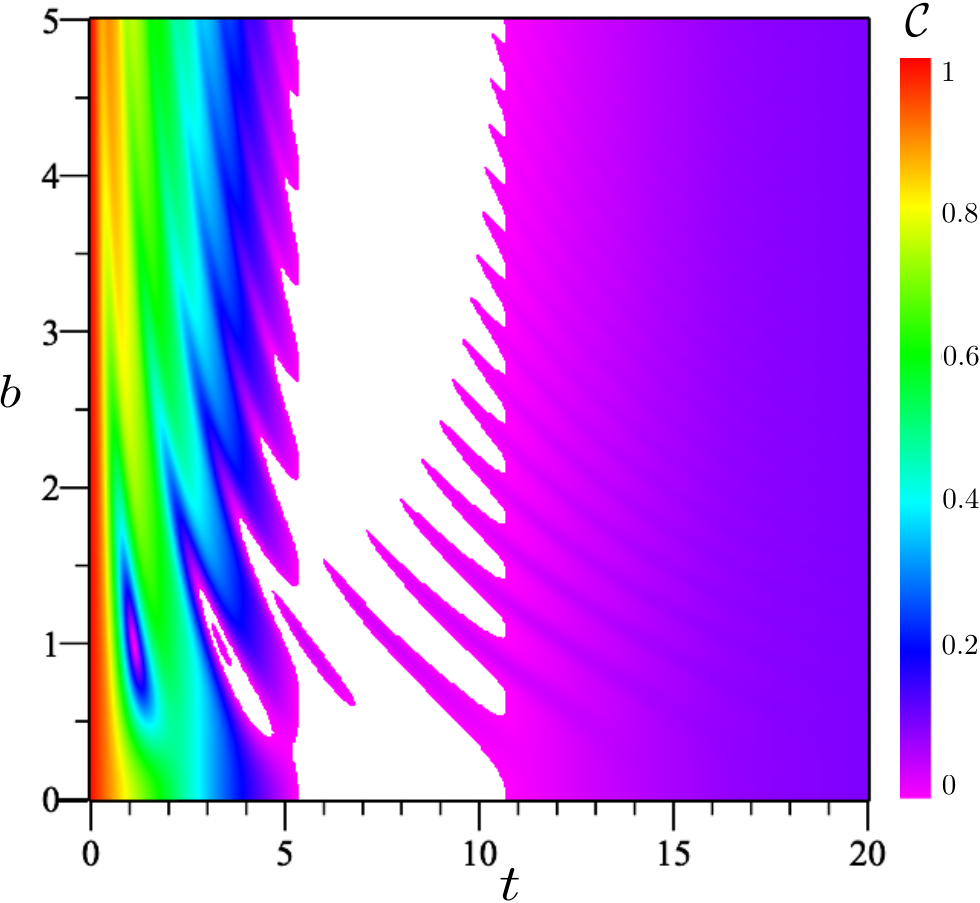}

\caption{\label{fig:dplot-C}(Color online) Density plot of concurrence $\mathcal{C}$ as a
function of time $t$  and magnetic field $b$.  The parameters are chosen as $J=2.0$, $J_{0}=1.0$, $B=0.2$, $\gamma=0.2$,
$\eta=0.2$ and $\theta=\pi/4$.}
\end{figure}

In Fig. \ref{fig:concurrence}, we conducted a more in-depth analysis of 
the dynamics of concurrence (Eq.\ref{eq:conc-def}) shows in the Figure \ref{fig:C1vst}(a), nothing 
that it consists of two distinct parts,  $\mathcal{C}_{1}$ and $\mathcal{C}_{2}$. Regarding $\mathcal{C}_{1}=2\mathrm{max}\{|\rho_{23}|-\sqrt{\rho_{11}\rho_{44}},0\}\ $ 
(red curve), we identified a competition between the initial state $|\Psi\rangle=\sin\theta|01\rangle+\cos\theta|10\rangle$ associate with $\rho_{23}$ and populations $\rho_{11}$ and $\rho_{44}$, which 
then abruptly disappear, expected to generate entanglement in the reservoirs. After a period of 
time without entanglement presence, a sudden birth of entanglement, $\mathcal{C}_{2}=2\mathrm{max}\{|\rho_{14}|-\sqrt{\rho_{22}\rho_{33}},0\}\ $(green curve), occurs 
as a result of interaction with the environment. Surprisingly, this quantum entanglement is generated by the state $\rho_{14}$, contrasting with the initial entanglement associated with the state $\rho_{23}$.
In Fig. \ref{fig:dplot-C}, we show the behavior of the concurrence $\mathcal{C}$ as a  function of time $t$ and $b$, with $\theta=\frac{\pi}{4}$. 
The figure reveals that near $b=1$, there is a narrow region characterized by abrupt events of entanglement sudden death and entanglement sudden birth around $t=3$. For $t>5$, a broad area emerges where entanglement dissipates and re-emerges recurrently, clearly indicating the dynamic interaction between the system and environment \cite{sol}. White regions indicate where the system is entirely unentangled.

\begin{figure}
\includegraphics[scale=0.35]{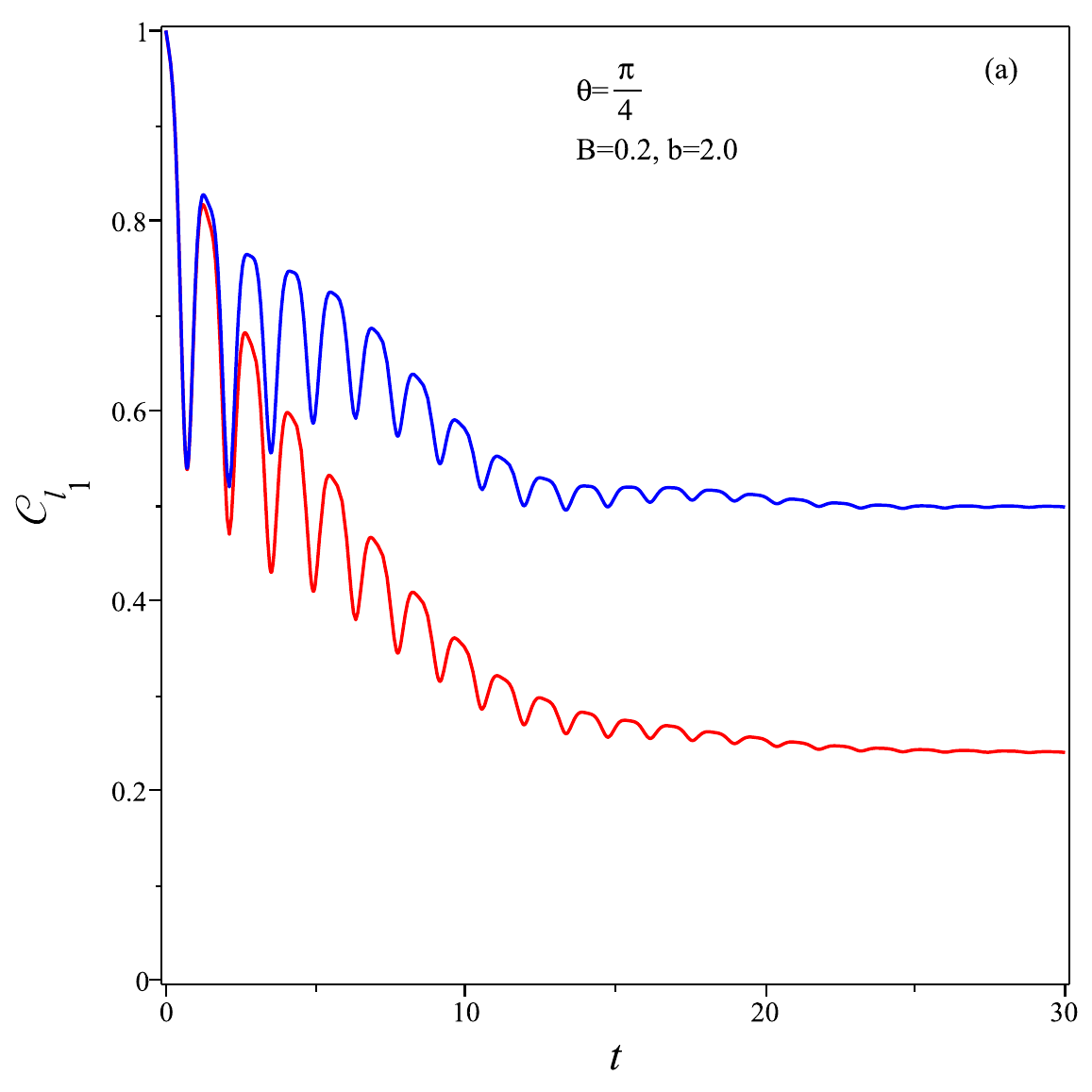}
\includegraphics[scale=0.35]{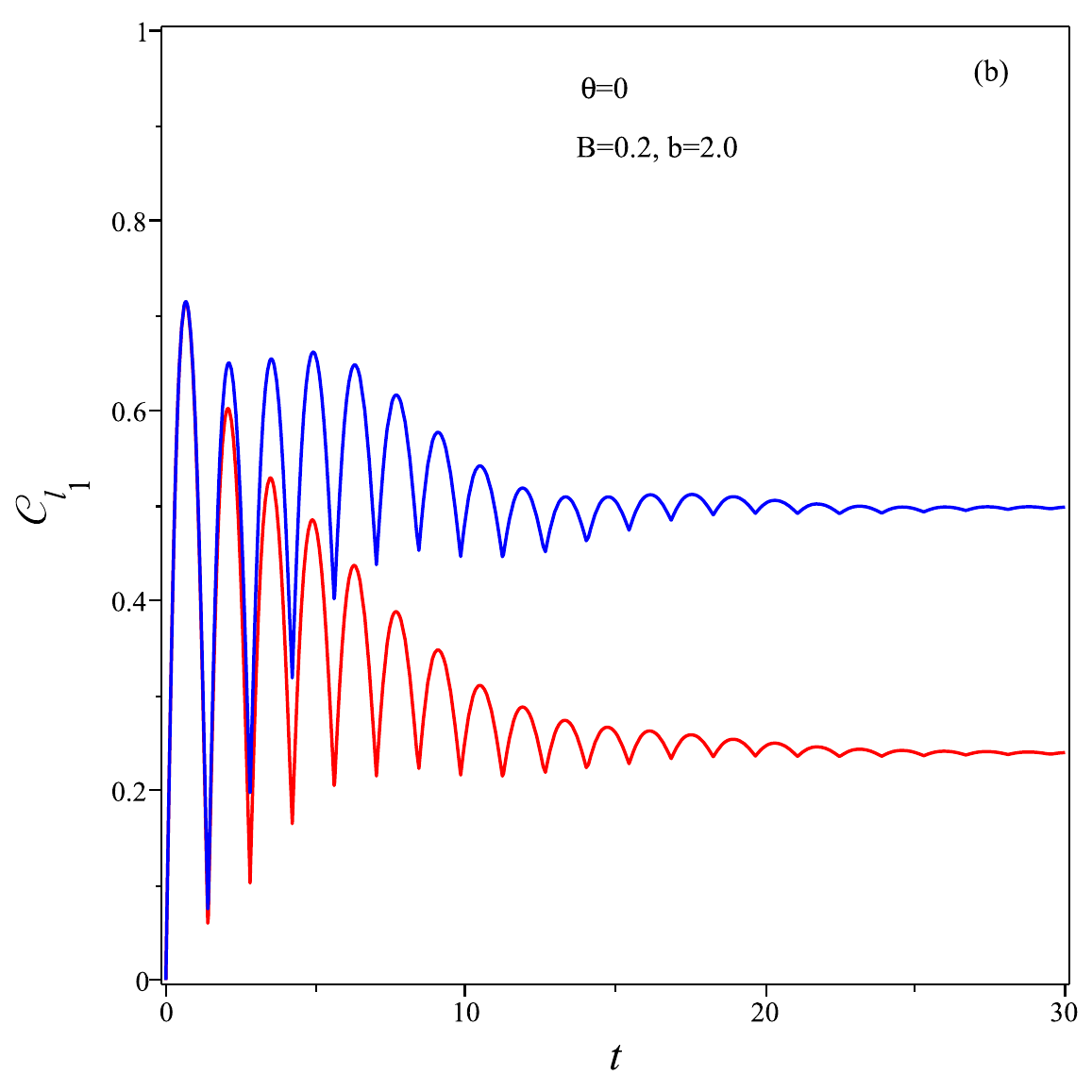}

\caption{\label{fig:Fig6}(Color online) The time evolution of quantum coherence $\mathcal{C}_{l_{1}}$ for  $B=0.2$, $b=2.0$. 
(a) $\theta=\frac{\pi}{4}$. (b) $\theta=0$. The other parameters are set to $J=2.0$, $J_{0}=1.0$, $\gamma=0.2$, $\eta=0.2$.}
\end{figure}

\begin{figure}
\includegraphics[scale=0.35]{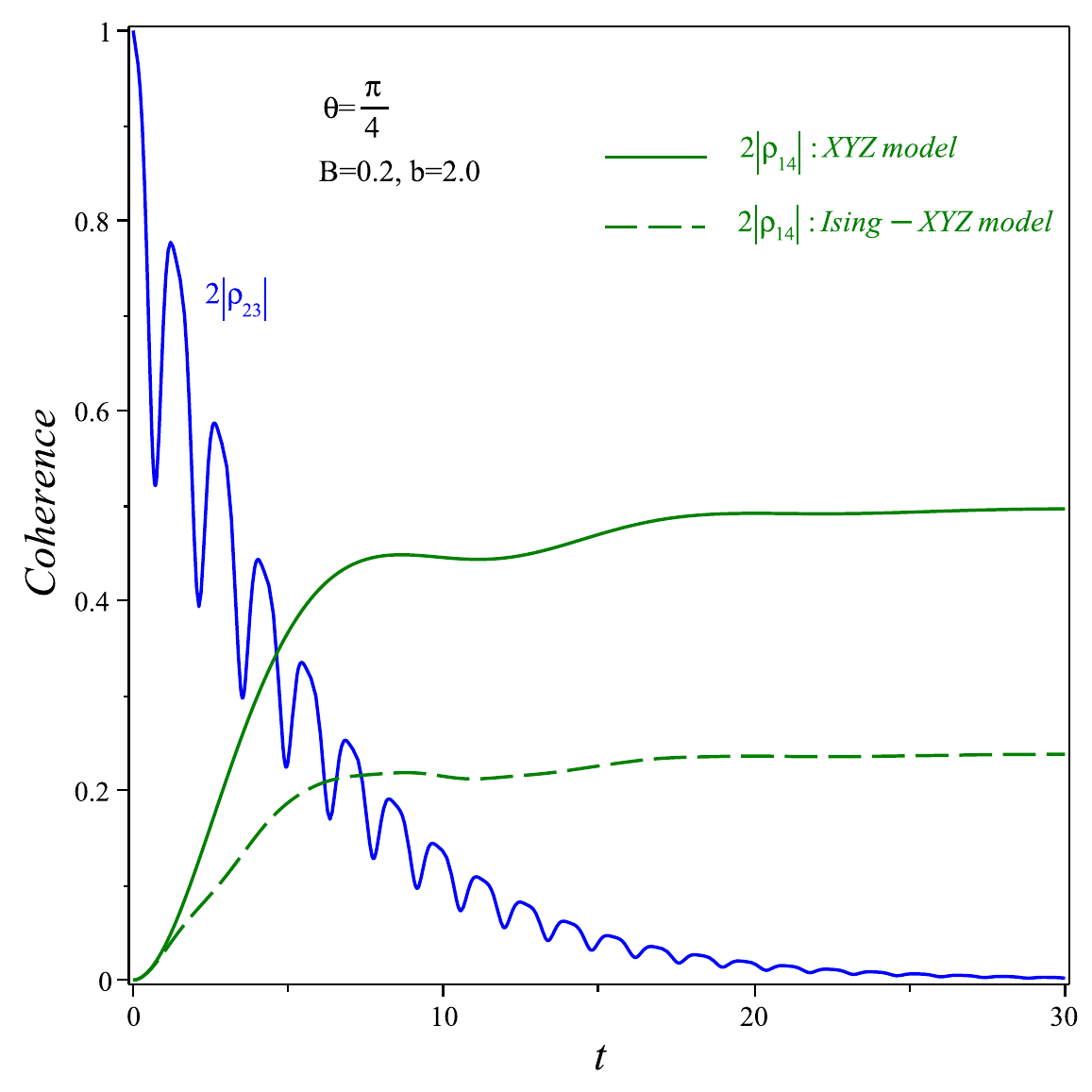}

\caption{\label{fig:Fig7}(Color online) The time evolution of quantum coherence $\mathcal{C}_{l_{1}}$ for  $B=0.2$, $b=2.0$.  $\theta=\frac{\pi}{4}$. The remaining parameters are fixed at $J=2.0$, $J_{0}=1.0$, $\gamma=0.2$, $\eta=0.2$.}
\end{figure}


\subsection{The $l_{1}$-norm of coherence}

To illustrate the behavior of quantum coherence in our model, the $l_{1}$-norm of coherence
$\mathcal{C}_{l_{1}}$ versus the time $t$ is plotted in Fig. \ref{fig:Fig6} for the magnetic fields $B=0.2$ and $b=2.0$, 
with fixed parameters $J=2.0$, $J_{0}=1.0$, $\gamma=0.2$, and $\eta=0.2$.
In Fig. \ref{fig:Fig6}(a), we plot the $l_{1}$-norm of coherence, $\mathcal{C}_{l_{1}}$, for the spin-1/2 Heisenberg $XYZ$ chain (blue curve) 
and the hybrid spin-1/2 Ising-$XYZ$ chain (red curve), a reduction in the $l_{1}$-norm of 
coherence $\mathcal{C}_{l_{1}}$  is observed when the Ising interaction is included in the model. Initially,  
the quantum coherence in the initial state is identical in both the Heisenberg model and the 
Ising-$XYZ$ hybrid model. However, as time evolves, we observe that the coherence decays more rapidly in the hybrid model compared to the original model. Subsequently, the $\mathcal{C}_{l_{1}}$ reaches the steady state after exhibiting an oscillatory behavior. In Fig. \ref{fig:Fig6}(b), we observe that on a short time scale for the initial unentangled state $|\Psi\rangle=|10\rangle$, the behavior of quantum coherence, captured by coherence $\mathcal{C}_{l_{1}}$, is indistinguishable between two models. However,  the norm-$l_{1}$-norm of coherence $\mathcal{C}_{l_{1}}$ in the original Heisenberg model proves to be more robust than in the hybrid Heisenberg Ising-$XYZ$ model, both during oscillatoring periods and when it reaches the steady state.

\begin{figure}
\includegraphics[scale=0.35]{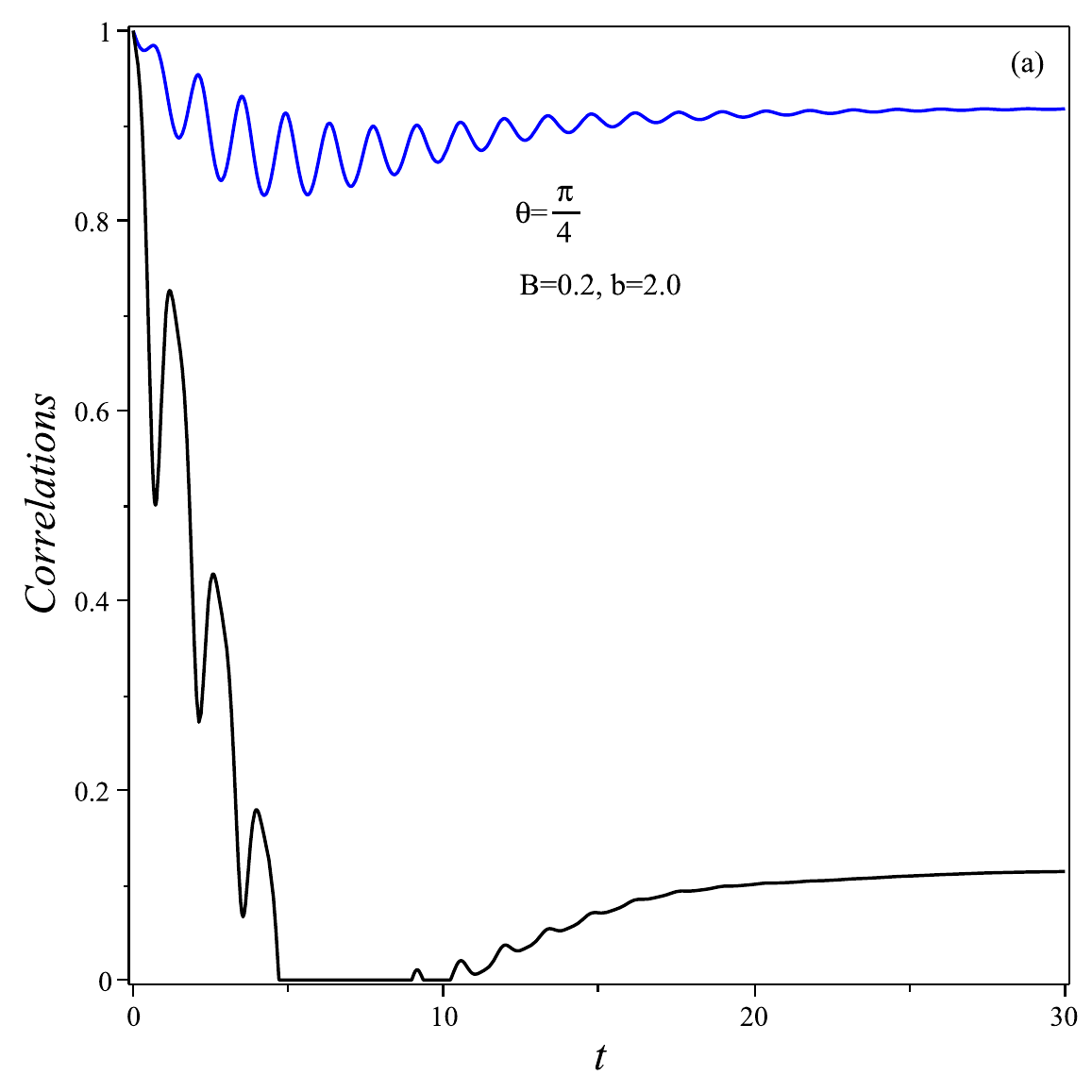}
\includegraphics[scale=0.35]{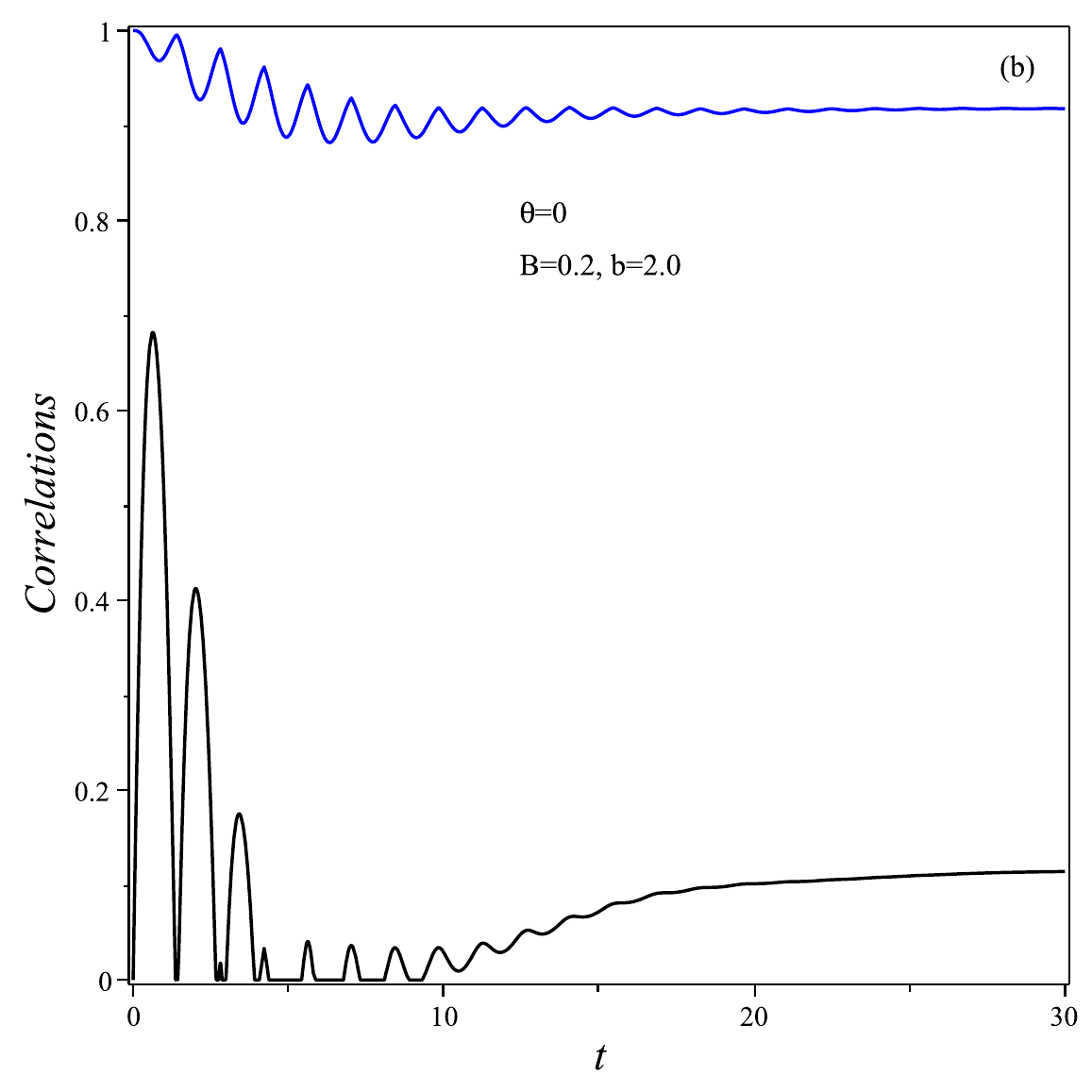}

\caption{\label{fig:LQFI}(Color online) Dynamics of the correlations as a
function of time $t$. Here the concurrence is black curve and LQFI is blue curve. (a) $B=0.2$, $b=2.0$, $\theta=\frac{\pi}{4}$. (b)
$B=0.2$, $b=2.0$, $\theta=0$. The remaining parameters are fixed at $J=2.0$, $J_{0}=1.0$, $\gamma=0.2$,
$\eta=0.2$.}
\end{figure}


In order to understand the behavior of quantum  coherence $\mathcal{C}_{l_{1}}$ in the examined models, we will perform a further analysis of the components of the $l_{1}$-norm depicted in Fig. \ref{fig:Fig6}(a). Thus, in  Fig. \ref{fig:Fig7}, we can observe that 
the loss of quantum coherence to the environment, represented by $2|\rho_{23}|$ (blue curve),  
is the same in both the original model and the Ising-$XYZ$ model. On the other hand, the 
quantum coherence associated with the state $2|\rho_{14}|$ returning from the environment is concentrated in the Heisenberg dimer (green curve) of the original model, while, in the 
case of the Ising-$XYZ$ model, the coherence redistributes between the Heisenberg dimer and the spins with Ising-type 
interaction (dashed green curve). Thus, the coherence originating from environment weakens 
due to the dispersion of a portion of it among spin-1/2 particles with Ising-type interactions.

\subsection{Local quantum Fisher information }

The local quantum Fisher information $(LQFI)$ is a crucial measure for investigating quantum correlation, particularly those of quantum discord type. In Fig. \ref{fig:LQFI}, we plot the $(LQFI)$ and concurrence as a function of time $t$, using the same parameters 
values of Fig \ref{fig:C1vst}. 
In Fig.  \ref{fig:LQFI}(a), we depicts the dynamics of the $LQFI$ and concurrence for the initial state $|\Psi\rangle=\frac{1}{\sqrt{2}}(|01\rangle+|10\rangle)$. It is evident that initially, the $LQFI$ aligns with the maximum concurrence of the initial state.  Over time, the concurrence diminishes due to interaction with the environment, while the $LQFI$ undergoes slight oscillations before a steady state.  In Fig.  \ref{fig:LQFI}(b), we observe that, despite the qubits being initially unentangled, the correlation induced by the environment causes a sudden birth of entanglement.  Following a series of sudden births and deaths of entanglement, the state eventually converges to a steady state. On the other hand, the system initially exhibits discord-like correlations quantified by the local quantum Fisher information. As time progresses,  the $LQFI$ displays a series of oscillations before converging to a steady state.

\begin{figure}
\includegraphics[scale=0.38]{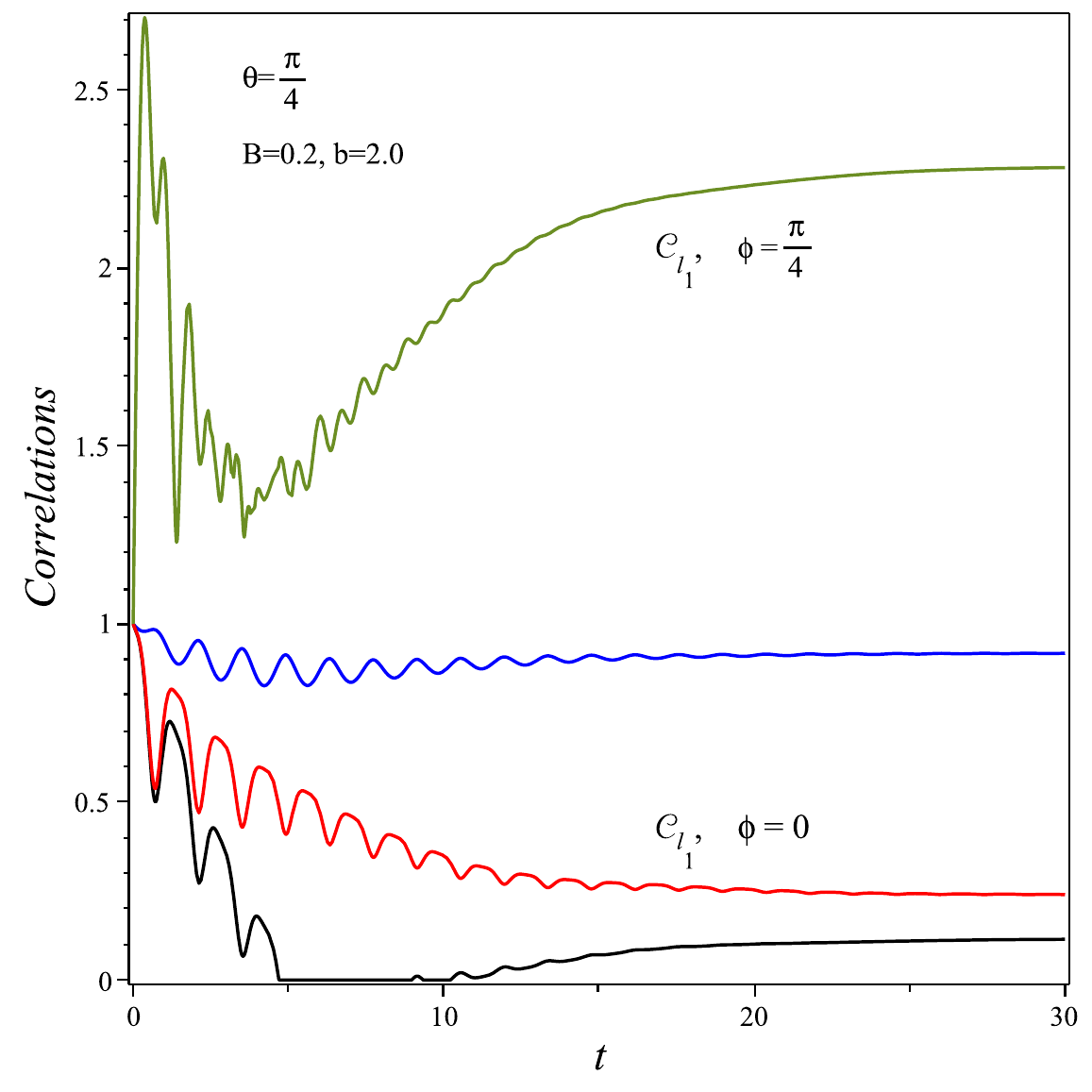}

\caption{\label{fig:geral}(Color online) Dynamics of quantum correlations as a
function of time $t$ for the parameter  $\theta=\frac{\pi}{4}$, $\phi=\frac{\pi}{4}$, $B=0.2$ and $b=2.0$. The other parameters are the same as those in Fig. \ref{fig:LQFI}.}
\end{figure}
Finally, to elucidate the variations in the behavior of the $l_{1}$-norm of coherence $\mathcal{C}_{l_{1}}$, it is important to note that the quantum coherence depends in the chosen basis. In Fig. \ref{fig:geral}, we depict the dynamics of quantum correlations ($\mathcal{C}$, $LQFI$,  $\mathcal{C}_{l_{1}}$) for the same parameters as in Fig. \ref{fig:LQFI}(a). Remarkably, the $LQFI$ is more robust than the coherence $\mathcal{C}_{l_{1}}$ (red curve) for $\phi=0$. This is because $\mathcal{C}_{l_{1}}$, for this choice of $\phi$, is composed exclusively of quantum correlations. Here, we choose $\phi=0$ in the transformation $\widetilde U$ (see Eq.\ref{u1}) to obtain $\mathcal{C}_{l_{1}}$. However, when selecting $\phi=\pi/4$, we observe a value  of   $\mathcal{C}_{l_{1}}$ (green curve) that exceeds both both $LQFI$ and $\mathcal{C}$. This suggests that, for this particular selection of $\phi$, quantum coherence incorporates quantum correlations, as well as local quantum coherences.

\section{Conclusions}

In this paper, we explore the  impact of decoherence on the dynamics of quantum coherence in the hybrid spin-1/2 Ising-$XYZ$ chain model with independent reservoirs at zero temperature. Furthermore, we contrast these effects with those observed in the spin-1/2 Heisenberg $XYZ$ model \cite{tao}. When comparing the coherence dynamics between the original model and the hybrid model, the research sheds light on the intricate interaction among quantum coherence, entanglement, and environmental decoherence in spin systems. The analysis provides insights into the evolution of quantum correlations over time and their convergence to steady states under the influence of environmental factors. Particularly, the Ising interaction in the hybrid model contributes to a slight increase in entanglement compared to the original model. In contrast, when analyzing quantum coherence, we observe a reduction in coherence in the hybrid model compared to the original model. This is due to the returning coherence from the environment being redistributed between the Heisenberg dimer and the Ising spins. In addition, it is important to note that the local quantum Fisher information has been found to be more robust than coherence in certain scenarios, highlighting the differential impact of environmental decoherence on various types of quantum correlations.
In conclusion, the study emphasizes the complex nature of quantum coherence dynamics in spin systems and the significant role of environmental decoherence in shaping quantum correlations.  The findings yield valuable insights into the behavior of entanglement, discord-like correlations, and coherence under environmental influences, opening opportunities for future research in quantum information processing and the dynamics of quantum correlations.

\section*{Acknowledgment}

O. Rojas, C. Filgueiras and M. Rojas thank CNPq, Capes and FAPEMIG
for partial financial support. H. L. Carrion acknowledges warm hospitality during his stay at Federal University of Lavras. M. Rojas acknowledges CNPq grant 317324/2021-7 and Fapemig Grant  APQ-02226-22.

\end{document}